\documentclass{dsj}
\setcitestyle{super,square,numbers,sort&compress}
\usepackage{array}
\usepackage{threeparttable} 
\usepackage{booktabs}
\usepackage{xcolor}
\usepackage{multirow}
\usepackage{ulem}
\usepackage{changes}
\usepackage{tabularx}
\bibpunct{(}{)}{;}{a}{,}{,}

\newcolumntype{P}[1]{>{\raggedright\arraybackslash}p{#1\textwidth}}
\newcolumntype{C}[1]{>{\centering\arraybackslash}p{#1\textwidth}}

\begin{document}

\footer{Footer Text}
\submitted{11 August 2024}

\title{How Generative AI Supports Human in Conceptual Design}

\author[1]{Liuqing Chen}
\author[1]{Yaxuan Song}
\author[1]{Jia Guo}
\author[1]{Lingyun Sun}
\author[2]{Peter Childs}
\author[2]{Yuan Yin\email{y.yin19@imperial.ac.uk}}

\address[1]{College of Computer science and technology, Zhejiang University, Zhejiang, China}
\address[2]{Dyson School of Design Engineering, Imperial College London, London, United Kingdom}

\maketitle

\begin{abstract}
Generative Artificial Intelligence (Generative AI) is a collection of AI technologies that can generate new information such as texts and images. With its strong capabilities, Generative AI has been actively studied in creative design processes. However, limited studies have explored the roles of humans and Generative AI in conceptual design processes, which leaves a gap for human-AI collaboration investigation. To address this gap, this study attempts to uncover the contributions of different Generative AI technologies in assisting humans in the conceptual design process. Novice designers were recruited to complete two design tasks in the condition of with or without the assistance of Generative AI. The results revealed that Generative AI primarily assists humans in the problem definition and idea generation stages, while the idea selection and evaluation stage remains predominantly human-led. Additionally, with the assistance of Generative AI, the idea selection and evaluation stages were further enhanced. Based on the findings, we discussed the role of Generative AI in human-AI collaboration and the implications for enhancing future conceptual design support with Generative AI’s assistance.

\end{abstract}

\section{Introduction}
Conceptual design, which translates design requirements into preliminary design solutions, is a crucial phase in the product design process (French, 1985). However, exploring problems and generating design solutions place high demands on designers' knowledge and reasoning abilities \citep{myrup2015conceptual}, reflecting the complex nature of conceptual design. Various design theories and methodologies have been proposed to help designers gain a more comprehensive understanding of the conceptual design process and to assist designers in developing creative ideas and solutions, such as TRIZ theory \citep{altshuller1999innovation}, FBS model \citep{gero2014function}, and C-K theory \citep{hatchuel2009ck}. These theories have made the representation of the conceptual design process more structured. However, the effective application of these methodologies still depends on designers' own knowledge and experience, which pose significant challenges for novice designers. With ongoing technological progress, some computational methods and tools have been proposed to alleviate novice designers' cognitive burden \citep{sarica2024innovation, cantamessa2020data}. For example, semantic networks \citep{luo2019computer, shi2017data} and case databases \citep{robles2009case, deldin2013asknature} have been established to support designers during the conceptual design stages. Although these tools provide inspiring stimuli, they do not offer corresponding solution suggestions for the specific design situation encountered. This means that designers still need to reason from the case domain to the problem domain to generate concrete solutions. 

Driven by technological advancements in machine learning, various Generative AI models including transformer \citep{vaswani2017attention}, diffusion models \citep{ho2020denoising}, and GANs (Generative Adversarial Networks) \citep{karras2019style} have demonstrated significant potential and powerful performances. Building on these technologies, applications such as ChatGPT, Stable diffusion, and Midjourney are making Generative AI more accessible and easy-to-use for consumers. Within the realm of design, text-to-text models and text-to-image models have attracted unique attention due to their ability to integrate seamlessly with the creative process and enhance the efficiency of design iterations. These models have become the most widely utilized generative techniques in research that combines Generative AI with design processes \citep{wu2024integrating}. Specifically, researchers have explored the application of text-to-text models in guiding the design process \citep{chen2024designfusion}, assisting with divergent and convergent thinking \citep{wang2023popblends}, and generating innovative solutions \citep{zhu2023generative}, holding great potential for creativity enhancement in the innovation process \citep{sarica2024innovation}. For text-to-image technologies, they help designers visualize their design ideas quickly and reduce the time and skill demand of manual sketching for human designers \citep{choi2024creativeconnect}. Also, text-to-image technologies can generate visual stimuli for design ideation based on user-input text prompt \citep{liu20233dall, wadinambiarachchi2024effects}.

Although researchers have recognized the importance of Generative AI in the conceptual design process, there is still a lack of empirical evidence for the effect of Generative AI in different stages of conceptual design. This gap may impede researchers from reflecting on and improving the developed collaborative tool designs. To fill the research gap, this study attempts to explore how Generative AI assists humans in conceptual design processes. Specifically, we recruited four groups of participants to finish two design tasks with (or without) the assistance of Generative AI (ChatGPT or Midjourney). We assessed human-AI collaboration in the conceptual design process through multiple dimensions, including the stages in which Generative AI helped designers, the stages led by humans, participants’ assessments of the Generative AI tool’s performance, expert ratings of the design outputs, and an prompt analysis of the strategies utilized by human designers during the four stages of conceptual design. We found that Generative AI primarily assists humans in the problem definition and idea generation stages, while the idea selection and evaluation stage remains predominantly human-led. Additionally, with the assistance of Generative AI, the idea selection and evaluation stage was further enhanced.

Our study provides an empirical contribution to the Generative AI-powered creativity support research by illustrating how Generative AI supports humans in conceptual design at a stage level. It further elaborates on Generative AI’s role in different stages across conceptual design. Further, we demonstrate implications for future conceptual design support under Generative AI’s help.

\section{Literature review}
\subsection{Conceptual design}

According to previous research, product design process can be divided into four phases: analysis of problem, conceptual design, embodiment of scheme, and detailing \citep{french1985conceptual}. Among these, conceptual design, which encompasses preliminary decision-making and design concepts generation, is regarded as the key part of the design process \citep{eppinger1995product}. A few conceptual design models have been proposed to explain the stages in conceptual design. For example, \cite{goodman2016designing} outlined that conceptual design process encompasses four stages: manage (deciding what actions to take next), explore (identifying needs), create (generating ideas), and evaluate (judging and testing the design concepts). \cite{Jasmine2020} delineated the design process into several distinct phases: establishing design requirements, assessing technology availability, sketching concepts and layouts, performing analysis and making trade-offs, optimizing revisions, and developing a preliminary design. Some researchers also promoted to apply the conventional design process model to conceptual design, such as the double diamond design process \citep{DesignCouncil2019}, which includes discover, define, develop, and deliver. Building on previous frameworks and considering the integral role of Generative AI, this study defines the conceptual design process as consisting of four stages: problem definition, idea generation, idea selection and evaluation, and idea evolution. This serves as the foundation for our experiment and underpins the research findings and conclusions presented in this study.

Although conceptual design is essential for design processes, it is challenging to obtain creative design ideas of high originality and novelty based on designers' own effort. Many computer-aided conceptual design support methods and tools have been proposed to offer creativity support to designers. For example, knowledge- or heuristics-based stimulation approaches can retrieval and mapping of source knowledge into the target design domain \citep{jiang2022data}. Some studies have attempted to utilize the information in patents, research papers, or encyclopedia data to construct semantic networks \citep{luo2019computer, sarica2020technet}. By computing the semantic distances between design goals and knowledge in database, these methods could offer design stimuli or knowledge to human designers. However, these stimuli-based methods still require designers' cross-domain reasoning to complete the final design concept adapting to the current problem scenario.

\subsection{Generative AI in conceptual design}
Driven by technological advancements in machine learning, such as Generative Adversarial Networks (GANs) \citep{goodfellow2014generative}, Variational Autoencoders (VAEs) \citep{kingma2014auto}, and transformers \citep{vaswani2017attention}, various Generative AI models including GPTs \citep{radford2018improving}, BERT \citep{kenton2019bert}, and StyleGAN \citep{karras2019style} have demonstrated significant potential and powerful performance. Among these, text-to-text and text-to-image models have garnered considerable attention in the field of conceptual design \citep{wu2024integrating}, and have sparked a series of studies on how to smoothly integrate these two types of models into existing workflows \citep{mahdavi2024ai, guo2024exploring}. 

Text-to-text tools, enhanced by Large Language Models (LLMs), such as ChatGPT, Llama, and BERT, can generate natural and fluent answers to comprehend user input and provide contextual solutions in natural language. In the conceptual design domain, Generative AI-based text generation has been applied to requirement extraction \citep{shahin2024harnessing}, creative ideation \citep{suh2024luminate}, solution generation \citep{chen2024triz} and so on. For our experiments, as the experiment was carried out in May 2023 to June 2023, we specifically chose GPT-3.5 due to its robust capabilities in generating coherent and contextually relevant text outputs, which has been widely applied in Generative AI-assisted design research \citep{chen2024designfusion, chen2024llm, chen2024beyond}. 

Some text-to-image models have also been used to aid designers in early concept development by providing visual references and multimodal stimuli \citep{kwon2023understanding}. These models, such as Midjourney, DALL-E, and Stable Diffusion, promote rapid exploration and iteration through visualization, enabling designers to better express their design concepts. Among these, Midjourney stands out both commercially and in terms of model performance, which has been widely adopted in human-AI collaboration research \citep{tan2024using, wadinambiarachchi2024effects, mahdavi2024ai} due to its impressive image generation quality and user-friendly features.

Although various work has been done to develop Generative AI-based design tools and methodologies, there is still a lack of empirical evidence for the effects of Generative AI in conceptual design processes. Thus, in this study, we adapt experimental methods from traditional design research to explore the influence of two representative Generative AI models (i.e. text-to-text and text-to-image models) on different conceptual design stages, contributing new empirical evidence to the design community. 

During the human-AI collaboration, Generative AI can assist human designers by generating concepts for selection, evaluation, and iteration. Additionally, the output from Generative AI can inspire designers to develop more innovative ideas as the information can often expand designers’ knowledge and exploration scope. These capabilities create unprecedented opportunities, particularly for novice designers, by significantly lowering the barriers to cross-disciplinary design and rapid visualization. Therefore, this study focuses on novice designers as research subjects to ensure more targeted research conclusions. 

\section{Experimental study}
\begin{figure}[htp]
    \centering
    \includegraphics[width=0.9\linewidth]{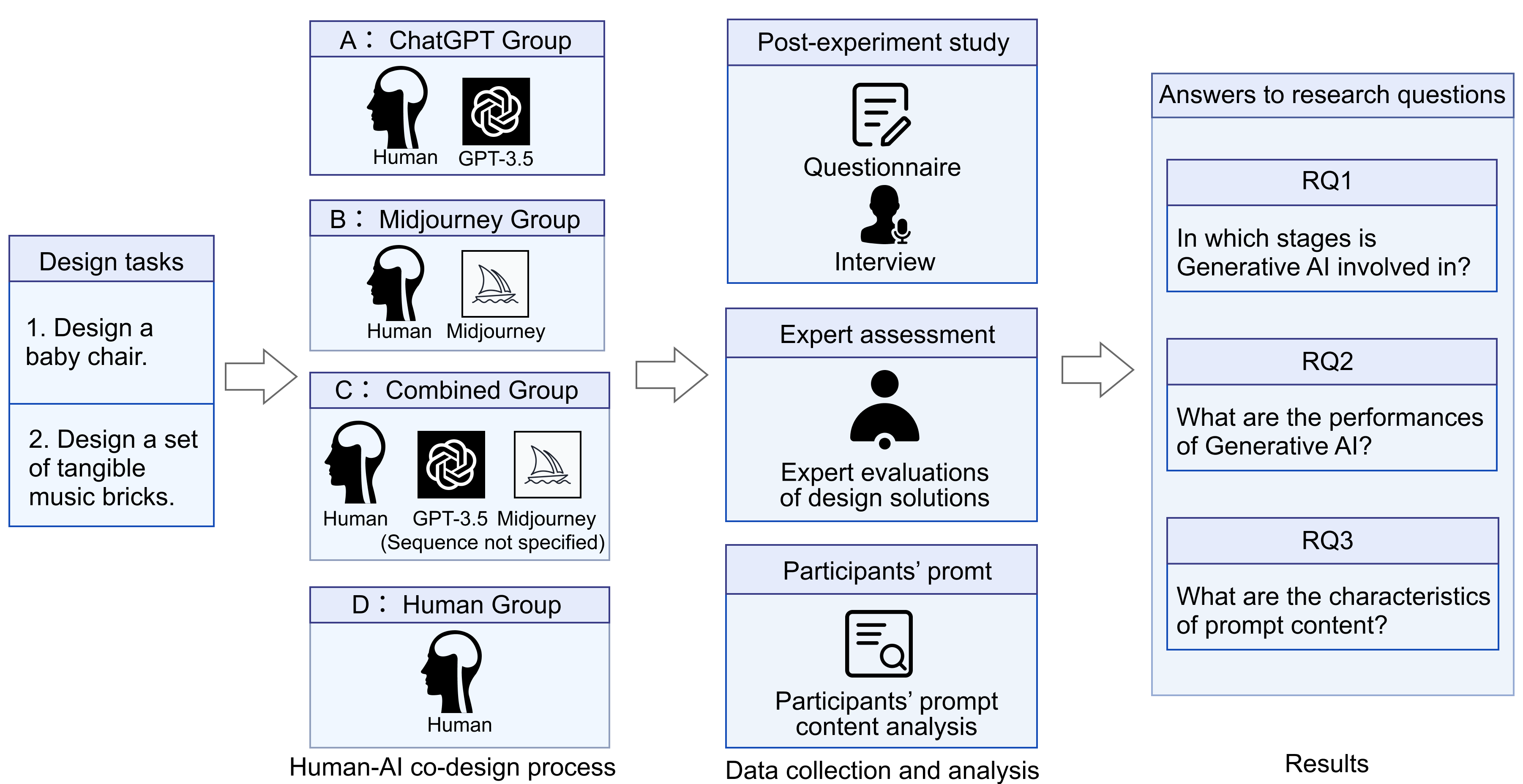}
    \caption{Representation of the experimental study procedure.}
    \label{fig:procedure}
    \vspace{-0.1in}
\end{figure}

In order to gain deeper insights into the human-AI collaboration paradigm, a human-AI co-design study was conducted. Midjourney (a text-to-image Generative AI model) and GPT-3.5 (a text-to-text Generative AI model) were selected as example Generative AIs. The selection of these general-purpose Generative AI tools, rather than design-specific alternatives, aligns with our research objective: to empirically investigate the role of Generative AI in conceptual design rather than to evaluate the existing design-specific tools. Furthermore, as identified in literature review, limited design-specific tools are designed to support the entire conceptual design process \citep{lee2024and}. Generative-purpose AI, therefore, is more suitable for achieving our research aim. The whole study procedure is shown in Figure~\ref{fig:procedure}. Primarily, we aimed to address three research questions through this experiment:

\begin{itemize}
  \item RQ1: In which stages is Generative AI involved in?
  \item RQ2: What are the performances of Generative AI?
  \item RQ3: What are the characteristics of prompt content?
\end{itemize}

\subsection{Participants}

We recruited participants through a university social networking site, with recruitment criteria requiring participants to be novice designers with less than four years of design learning experience. Additionally, it was required that participants have prior experience with ChatGPT and Midjourney. This can enable participants to focus more on design tasks rather than adapting to new Generative AI tools during the experiment. There were no restrictions for major background. After the screening process, a total of twenty participants (thirteen females, seven males, aged 18-26, SD=2.7) who met these criteria were selected. Each participant was paid \$10 per hour, and the average time for completion was about 70 minutes. 

\subsection{Procedure}

Participants were randomly divided into four groups, which are ChatGPT Group, Midjourney Group, Combined Group, and Human Group. It is important to note that in the Combined Group, participants can use both ChatGPT and Midjourney freely for both tasks, without limitations on orders. Before the experiment started, we conducted a 20-minute training session that covered the basics of conceptual design procedures and how to utilize GPT-3.5 and Midjourney to generate conceptual designs. Then, the experimental procedure and two conceptual design tasks were introduced to each participant. Two distinct design tasks were selected to mitigate potential biases, such as participants' potential expertise in a single task-related area \citep{hu2018effects}. The first task required participants to design a baby chair in 20 minutes and the second task involved designing a tangible music bricks in 20 minutes. The selection of the two design tasks was based on two considerations: firstly, the design ideation of the task could be conveyed through shape and external structure instead of intricate details related to internal structure. Secondly, the conceptual design task should involve objects that participants are familiar with but not commonplace to ensure participants can complete the design tasks while having room for creative divergence. After introducing the experiment tasks and addressing the participants' questions about the procedure, the formal experiment began.

During the experiment, there are some specific requirements for each group:

\begin{itemize}
  \item ChatGPT Group: Participants were asked to use ChatGPT to complete both of the design tasks.
  \item Midjourney Group: Participants were asked to use Midjourney to complete both of the design tasks.
  \item Combined Group: Participants were asked to use ChatGPT and Midjourney to complete both of the design tasks. The sequence of tool usage was not predetermined, allowing participants to choose the order based on their preferences.
    \item Human Group: Participants were asked to finish both of the tasks on their own (without the assist from Generative AI).
\end{itemize}

Finally, each participant was required to create an image for each task that illustrates the product’s design features, accompanied by essential text descriptions to clarify the design.

After the two tasks, each participant was invited to fill out a questionnaire. The questionnaire encompasses a 7-point Likert scale across five criteria regarding the evaluation of Generative AI’s performance (Participants in the Human Group were invited to evaluate their performance to serve as a baseline for comparing the performance of Generative AI). The questionnaire also explored the participants' reflections about which stages Generative AI helped with, and which stages are human-led (see details in Section~\ref{section_3.3.1}). In the semi-structured interview, we discussed the questionnaire results with participants and their attitudes, evaluations, and suggestions regarding the Generative AI-assisted human-AI collaboration. Each interview lasted around thirty minutes. All study procedures conformed to the Institutional Review Board (IRB) guidelines on human subject study.

\subsection{Data collection and analysis}
\subsubsection{Participants’ assessment of Generative AI tools performances}
\label{section_3.3.1}
After completing the experimental tasks, a performance assessment questionnaire was distributed to each participant. In the questionnaire, participants needed to evaluate the performance of Generative AI tools which they used on six criteria: speed, subject, diversity, novelty, triggering more ideas, and requirement satisfaction. The performance evaluation focused on the overall design process. These criteria were selected as they effectively reflect the impact of Generative AI in conceptual design. To be specific, the criteria for diversity and novelty were inspired by traditional conceptual design evaluations \citep{shah2003metrics}, while the other criteria (trigger more ideas, requirement satisfaction, speed, and subject) were specifically formulated based on key factors in Generative AI-assisted conceptual design and related to the objectives of this study. Detailed explanations of the six criteria are described in Table 1. Additionally, for the Combined Group, the questions were designed to evaluate ChatGPT and Midjourney separately. For the Human Group, participants were asked to assess their own performance. This approach allowed us to collect firsthand feedback from human designers and gain insights into their modes of collaboration with AI. 

\begin{table}[htp]
\centering
\caption{Participants' evaluation criteria.}
\label{tab: criteria_participants}
\resizebox{1.0\textwidth}{!}{%
\vspace{0.05in}
\begin{tabular}{P{0.3}P{0.7}} 

\toprule
\textbf{Criterion} & \textbf{Description} \\
\midrule
Speed & Measures how quickly the AI can complete conceptual design tasks, emphasizing efficiency in the design process.  \\
\midrule
Subject & Assesses whether the AI-generated outputs align with logistical requirements, ensuring they meet specific project needs. \\
\midrule
Diversity & Evaluates the variety of ideas produced by the AI, reflecting its capability to explore a wide range of creative solutions. \\
\midrule
Novelty & Rates the uniqueness and originality of the AI's outputs, crucial for innovation in design. \\
\midrule
Trigger more ideas & Looks at the AI’s ability to inspire further creativity among human collaborators, enhancing the ideation process. \\
\midrule
Requirement satisfaction & Checks if the AI can accurately interpret prompts and produce results that fulfill user requirements, showing adaptability and responsiveness.  \\
\bottomrule

\end{tabular}
}
    \begin{tablenotes}
    \footnotesize
    \item[] \textbf{Notes:}
    Participants in the Human Group were invited to evaluate their own performance to serve as a baseline.
    \end{tablenotes}
    \label{tab:expert_rate}
\end{table}

\subsubsection{Expert ratings}
\label{subsubsection:expert_ratings}
Five professional designers (three males, two females, aged 25-29), who have more than five-year design experience, were recruited as experts to evaluate the conceptual design solutions created by the participants in the four groups. During the evaluation, the forty design solutions were randomly displayed. For each solution, assessors were first introduced which task (Task 1 or Task 2) the solution was from. Then, assessors were asked to assess the solutions using 7-Likert scale (1: The performance is really poor; 2: The performance is poor; 3: The performance is below average; 4: The performance is average; 5: The performance is above average; 6: The performance is very good; 7: The performance is perfect). The experts assessed the design solutions based on five primary criteria: (1) Novelty: whether the design introduces new ideas or approaches that are not commonly found in similar products; (2) Feasibility: whether the design can be realistically implemented; (3) Usability: whether potential users can easily and effectively use the product to achieve their goals; (4) Functional diversity: the range of functions that the design can perform; and (5) Cost: the overall expenses involved in manufacturing, distributing, and maintaining the product over its lifespan (high cost score means poor performance). The assessment process lasted around 30 minutes.

\subsubsection{Generative AI's helpful stages in conceptual design}
As this study aims to characterize human-AI collaboration in conceptual design at a stage level, we defined ``actual-helping stages" as the stages where participants reported completing tasks with the assistance of Generative AI. To explore this, participants needed to fill the post-experiment questionnaire to report the stages in which Generative AI actually helped them. Additionally, participants needed to report which stages were led by humans, implying that human designers completed most of the work independently. These two questions were presented in the form of multiple-choice questions, allowing participants to select all the stages they felt were applicable. The questionnaire for the Combined Group is detailed in Appendix A as an example. 

\subsubsection{Participants' prompt}
\label{subsubsection:prompt_analysis}
All text inputs by participants to communicate with Generative AI during their conceptual design process were collected. In total, we gathered 114 prompts across the ChatGPT Group, Midjourney Group, and Combined Group for Tasks 1 and 2, averaging 3.8 prompts per participant per task. For the data analysis process,  we first categorized participants' prompt into one of the four stages of conceptual design. Initially, a random sample of three participants' prompt from each group was selected, and two researchers independently categorized these samples to develop a preliminary understanding. After discussing their individual classifications and explanations, they reached a consensus, which facilitated the finalization of a comprehensive codebook, detailed in Appendix~B. After establishing the codebook, the two researchers independently coded the prompts for the remaining two participants' prompt in each group, achieving an inter-rater reliability of $\kappa = 0.74$, indicating a strong agreement between the coders. Ultimately, in the four stages analyzed, there were 31, 14, 10, and 9 prompts identified with ChatGPT, and 13, 20, 5, and 12 prompts identified with Midjourney, respectively.

We then moved to the next phase of our analysis, which involved systematically summarizing the strategies ChatGPT and Midjourney assisted human designers with during each stage of the conceptual design process. Specifically, the same two researchers independently reviewed the categorized prompts to identify the assistance strategies provided by ChatGPT and Midjourney for each stage. Discussions were frequently made to resolve any discrepancies. Specifically, we applied the affinity diagramming method to aggregate and analyze the topics reflected in participants' prompt \citep{holtzblatt1997contextual}. In this process, two researchers placed the original prompt contexts onto sticky notes, grouped them, and iteratively labeled each group with descriptors to elucidate their shared themes. The summarized strategies, along with corresponding examples, are presented in Section~\ref{subsec_RQ3_result}.



\subsubsection{Post-experiment interview}
We conducted one-to-one interviews after the participants finished the two design tasks and questionnaire to gain deeper insights into how novice designers collaborate with Generative AI during the conceptual design process. The interview questions were tailored based on the participants' questionnaire responses and the design solutions they completed, focusing on the following aspects:

(1) Why and how did you use ChatGPT/Midjourney during the [specific design stage]? 

(2) In the questionnaire, you rated [specific criterion] with [specific score]. Why did you give this rating?

(3) In the conceptual design's human-AI collaboration, you mentioned that the [specific design stage] should be human-led. Why do you think so? Furthermore, you indicated that [specific design stage] requires collaboration between humans and Generative AI. Could you please explain this in detail?

(4) What are your other feelings and views about the collaboration between humans and Generative AI in conceptual design?

\section{Results}
In this section, the three research questions are systematically answered based on the analysis from the collected data. First, we answer the question of which stages that Generative AI is involved in in RQ1. Second, we explore the performances of Generative AI in the conceptual design process both from participants perspective and expert ratings in RQ2. Third, in RQ3, we delve into the prompt analysis results from human designers.

\subsection{RQ1: In which stages is Generative AI involved in?}
\label{result_RQ1}
\begin{figure*}[htp]
    \centering
    \includegraphics[width=0.9\linewidth]{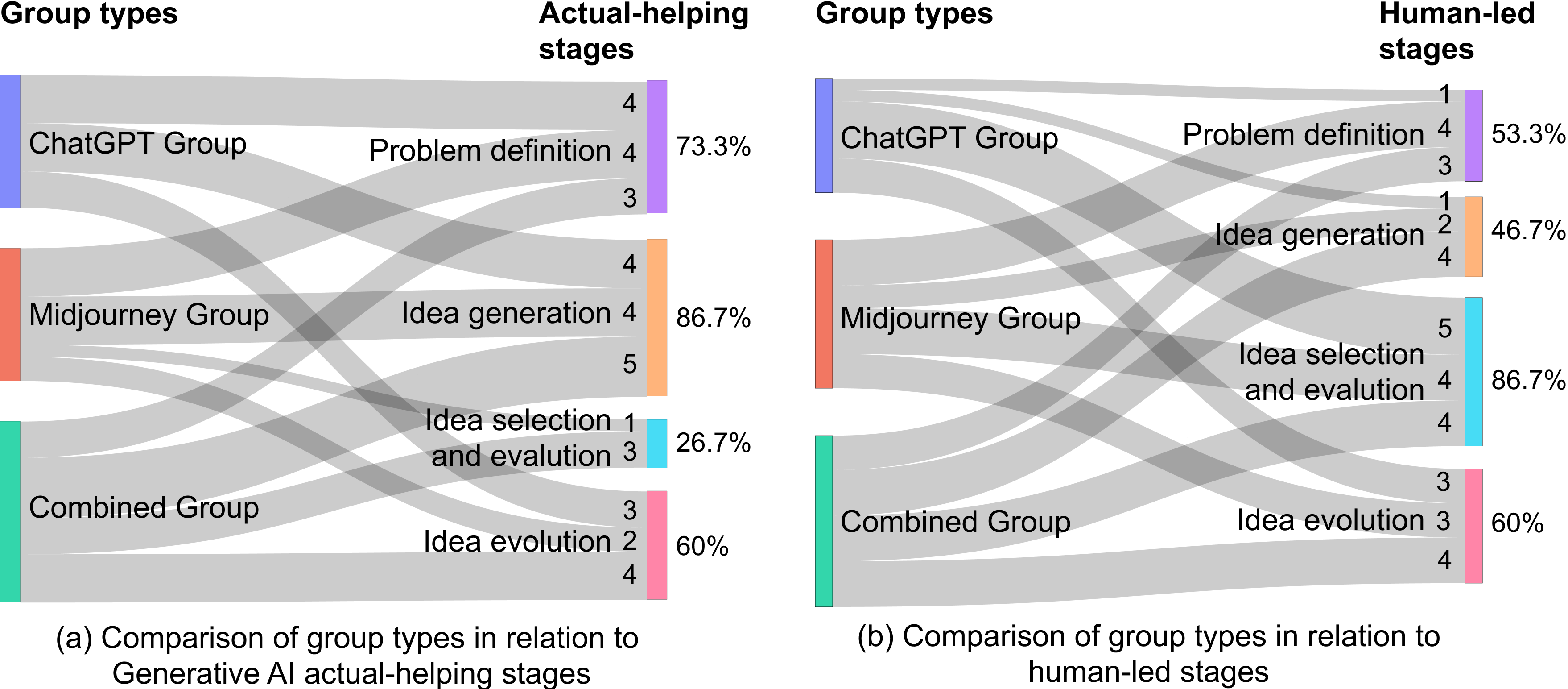}
    \caption{Horizontal Sankey diagrams representing (a) the comparison of group types in relation to Generative AI’s helping stages and (b) the comparison of group types in relation to human-led stages. (Percentages in the figure represent the proportion of responses among the fifteen participants in Generative AI-assisted groups.)}
    \label{fig:sankey}
    \vspace{-0.1in}
\end{figure*}

We initially identified the stages where Generative AI assisted designers and those perceived as human-led. This analysis includes data from the ChatGPT Group, Midjourney Group, and Combined Group. Figure~\ref{fig:sankey} illustrated Generative AI’s helping stages and human-led stages in conceptual design processes by two Sankey diagrams. The percentages represent the proportion of responses among the fifteen participants. Figure~\ref{fig:sankey} (a) demonstrates that Generative AI predominantly supported humans during the idea generation stage, problem definition stage, and idea evolution stage. Respectively, 86.7\%, 73.3\%, and 60\% of participants recognized the assistance of Generative AI in these stages. This indicates that text-to-text and text-to-image Generative AI tools are particularly effective in initiating and nurturing early-stage design thinking, where conceptual blending and broad brainstorming are crucial \citep{wang2023popblends}. Figure~\ref{fig:sankey} (b) reveals that the idea selection and evaluation stage (86.7\%), as well as the idea evolution stage (60\%), are predominantly perceived as human-led. This could imply that human judgment remains essential when it comes to evaluating and making final decisions on these ideas. 

Overall, while Generative AI primarily supports the early stages of conceptual design, such as problem definition and idea generation, it still relies on human-led processes during the evaluation phase, highlighting the complementary roles of Generative AI and human expertise play in the design process.

\subsection{RQ2: What are the performances of Generative AI?}
This subsection synthesizes the assessments of Generative AI's performance by participants with expert evaluations of the final design solutions. By integrating these perspectives, we aim to provide a multidimensional understanding of how AI tools contribute to and influence the conceptual design process. As for data analysis, ANOVA analysis was employed when the data followed a normal distribution. For a non-normal distribution, the Kruskal-Wallis H test, a non-parametric test, was utilized to detect significant differences between the four groups.

Table~\ref{tab:participant_evaluation_seven} presents the assessments of five evaluation criteria by participants during the experimental process. Through the statistical results, some insights could be gleaned regarding model characteristics and human-AI collaboration. Notably, ChatGPT Group excelled in speed and requirement satisfaction. The superior speed performance can be attributed to ChatGPT's text-to-text model features, which allow for faster generation and real-time progress tracking. In contrast, Midjourney applies an iterative refinement process, starting with an initial visual pattern that progressively evolves into cleaner outputs through multiple enhancement steps. In addition, the requirement satisfaction scores were lower when participants used Midjourney, which means that instructions were not always effectively reflected in the final images produced. This reflects challenges in controlling Midjourney, as participants frequently reported that while they could manipulate shape design, the details did not align with their intentions (\textit{P2-Midjourney: ``When I input some commands, the generated images could generally shape the overall appearance, but the finer details did not align with my intended design specifications."}). In contrast, Midjourney achieved the highest scores in terms of subject, diversity, novelty, and triggering more ideas, likely benefiting from its visual representation which offers more direct stimuli. When comparing scores with and without Generative AI tools’ help, ChatGPT Group achieved a lower score than the Human Group in subject, primarily because the Generative AI's outputs were probabilistic rather than fact-based \citep{brown2020language}. 

\begin{table}[htp]
\centering
\caption{Average scores and standard deviations of participants’ evaluation of different Generative AI of each group in experimental design.}
\label{tab:participant_evaluation_seven}
\resizebox{0.9\textwidth}{!}{%
\begin{tabular}{P{0.30}*{4}{C{0.22}}}
\toprule
Criterion & ChatGPT Group & Midjourney Group & Combined Group & Human Group* \\
\midrule
Speed & \textbf{6.0} (0.71) & 5.0 (1.00) & 4.6 (1.14) & N/A \\
Subject & \underline{4.6} (1.14) & \textbf{5.6} (1.14) & 5.2 (0.84) & 4.8 (0.84) \\
Diversity & 3.8 (1.48) & \textbf{5.2} (1.48) & 4.6 (1.14) & 3.4 (0.55) \\
Novelty & 3.8 (0.89) & \textbf{5.4} (1.14) & 4.0 (1.22) & 3.6 (0.55) \\
Trigger more ideas & 4.8 (1.30) & \textbf{5.6} (0.89) & 4.6 (1.14) & 4.6 (1.14) \\
Requirement satisfaction & \textbf{5.0} (1.87) & 4.8 (0.84) & 3.8 (0.84) & N/A \\
\bottomrule
\end{tabular}
}
\begin{tablenotes}
    \footnotesize
    \item[] \textbf{Notes:}
    \item 1. * indicates our baseline condition.
    \item 2. \textbf{bold} indicates the best performance among four groups.
    \item 3. \underline{underline} indicates performance worse than Human Group.
\end{tablenotes}
\end{table}

The statistical assessment of expert ratings, detailed in Table~\ref{tab:expert_rating_results}, shows that the Midjourney Group achieved the highest overall score with a mean of 4.34 and a standard deviation of 0.75. Significant differences were observed in the metrics of novelty $(p<0.01^{**})$, cost $(p<0.01^{**})$, and overall performance $(p<0.01^{**})$. Subsequent post hoc tests utilizing the Bonferroni correction revealed significant differences in novelty between the Human Group and the Midjourney Group $(p<0.01^{**})$, Human Group and the Combined Group $(p<0.01^{**})$. Additionally, the ChatGPT Group's mean score of 3.98 also surpassed that of the Human Group's 3.20, indicating that Generative AI tools could broaden the range of design options and introduce unique visual examples that enhance creativity. In addition, significant differences were found in the cost metric between the Human Group and both the ChatGPT Group $(p<0.01^{**})$ and the Midjourney Group $(p<0.01^{**})$. Notably, a higher cost score implies poorer performance, suggesting that the use of Generative AI may increase the complexity of design ideas, according to the definition of cost in Section~\ref{subsubsection:expert_ratings}. The same pattern was observed in overall scores, where both the ChatGPT Group and the Midjourney Group outperformed the Human Group, with $p=0.02^{*}$ and $p=0.01^{*}$ respectively. The Combined Group's mean score of 4.10 also surpassed the Human Group's 3.59. These results reflect that, compared to Human Group, human designers with Generative AI tools consistently achieved higher scores regarding expert ratings. The average Cohen’s kappa among five assessors was 0.66 (with detailed results shown in Appendix~C), which indicates an acceptable level of consistency.

\begin{table}[htp]
\centering
\caption{Expert rating results combined two tasks.}
\label{tab:expert_rating_results}
\resizebox{1.0\textwidth}{!}{%
\begin{tabular}{p{3cm}cccccc}
\toprule
\multirow{2}{*}{\textbf{Group}} & \multicolumn{6}{c}{Score (SD)} \\
\cline{2-7}
 & Novelty & Feasibility & Usability & Functional diversity & Cost & Overall \\
\midrule
ChatGPT Group & 3.98 (1.29) & 4.56 (0.83) & \textbf{4.24} (1.12) & \textbf{3.88} (1.01) & 4.52 (0.91) & 4.24 (0.60) \\
Midjourney Group & 4.60 (1.23) & \textbf{4.62} (1.09) & 4.10 (1.04) & 3.60 (1.31) & 4.76 (1.23) & \textbf{4.34} (0.75) \\
Combined Group & \textbf{4.90} (1.31) & 4.24 (0.89) & 3.80 (0.78) & 3.38 (0.96) & 4.18 (1.25) & 4.10 (0.66) \\
Human Group & 3.20 (1.17) & 4.08 (0.77) & 3.70 (0.92) & 3.44 (1.08) & \textbf{3.52} (0.91) & 3.59 (0.71) \\
\midrule
P-value & 0.000 \textsuperscript{**} & 0.084 & 0.228 & 0.422  & 0.001 \textsuperscript{**} & 0.005\textsuperscript{**} \\
\bottomrule
\end{tabular}
}
\begin{tablenotes}
    \footnotesize
    \item[] \textbf{Notes:}
    \item 1. ** denotes \(p < 0.01\) and * denotes \(p < 0.05\).
    \item 2. Bolded scores indicate the best performance among the four groups.
\end{tablenotes}
\end{table}


\subsection{RQ3: What are the characteristics of prompt content?}
\label{subsec_RQ3_result}
After completing the coding process described in Section~\ref{subsubsection:prompt_analysis}, a summary of the strategies employed by human designers, alongside relevant examples, is presented in Appendix~D due to space constraints. Subsequent sections will discuss these strategies as employed across the four stages of conceptual design: problem definition, idea generation, idea selection and evaluation, and idea evolution.

\subsubsection{Problem definition stage:}
For ChatGPT, the highest number of prompts was identified in this stage (31/64 = 48.4\%), with five strategies summarized. Firstly, it aided in \textbf{identifying the target audience} by providing demographic data, illustrated by a response detailing the age range for baby chair usage \textit{(P5-ChatGPT Group)}. Secondly, ChatGPT assisted in \textbf{user needs analysis}, as shown by an inquiry about parents’ needs for baby seats \textit{(P1-ChatGPT Group)}, gathering insights that traditionally depends on extensive research \citep{french1985conceptual}. Thirdly, participants utilized ChatGPT to offer insights into \textbf{existing products and their market status}, which streamlined the research phase by providing comprehensive and organized answers, exemplified by a prompt about current music visualization tools \textit{(P1-Combined Group)}. Fourthly, \textbf{functionality considerations} were explored through inquiries about necessary features for a baby seat \textit{(P5-ChatGPT Group)}. Finally, participants also employed ChatGPT to investigates suitable \textbf{materials} that meet both the functional and aesthetic needs of the product \textit{(P4-ChatGPT Group)}.

For Midjourney, it enhanced the problem definition stage by \textbf{facilitating the exploration of various design intents}. It enabled designers to experiment with different aesthetic and functional styles visually, thus aiding the formulation of their own design concepts. A representative prompt is demonstrated by the P4-Midjourney Group \textit{``baby chair, cute, bright colors"}, which served as the initial input to explore a variety of design elements from an initial vague design direction.

\subsubsection{Idea generation stage:}
ChatGPT facilitated idea generation in two distinct ways, differentiated by whether designers had initial design elements. These methods are identified as \textbf{``key design points synthesis"} and \textbf{``intuitive idea generation"}. For instance, with specific design elements in mind, the P4-ChatGPT Group formulated a prompt to \textit{``Design a baby chair that combines growth adaptability, non-toxic materials, and music."} Meanwhile, the tool could also generate original designs spontaneously without specific directions from designers, as seen in the P1-Combined Group’s prompt: \textit{``Design an innovative baby chair."}

For Midjourney, the highest number of prompts was identified in this stage (20/50 = 40.0\%), enhancing the idea generation stage in a relatively straightforward way by efficiently \textbf{transforms design ideas into visual representations}. This quick visualization saved much time and effort for manual sketching, which also facilitated the following idea selection and evaluation process.

\subsubsection{Idea selection and evaluation stage:}
In the idea selection and evaluation stage, ChatGPT enhanced the design process by providing two essential types of support: \textbf{creativity evaluation} and \textbf{feasibility assessment}. Creativity evaluation primarily focuses on the novelty of design concepts. It involves ChatGPT aiding designers by highlighting innovative elements and suggesting areas for enhancement \textit{(P4-Combined Group)}. On the other hand, feasibility assessment concentrated on evaluating the practicality of the proposed concepts \textit{(P1-ChatGPT Group)}. 

Meanwhile, Midjourney contributes by enabling the \textbf{creation of multiple design variants}. This feature facilitates the comparison and selection among different design concepts. For example, P3-Midjourney Group, who re-entered a prompt from an earlier step to generate more visualizations, explored various adaptations of a specific design idea. This process underscores Midjourney’s capability to quickly adapt and visualize numerous iterations.

\subsubsection{Idea evolution stage:}
In the idea evolution stage, designers primarily leveraged ChatGPT to enhance the design process in two ways: \textbf{refining design elements} and \textbf{facilitating concept iteration}. For the refinement of design elements, designers employed ChatGPT to improve and elaborate on the proposed solution's details. An example is the P3-ChatGPT Group's use of ChatGPT to refine a children's seat design by integrating more comfortable materials, as illustrated in the prompt: \textit{``Refine the integration of fabric and Lego to optimize comfort and functionality in the children's seat design."} In terms of concept iteration, designers revisited and revised their initial design directions. The revision process is exemplified by the P5-ChatGPT Group’s request to \textit{``Propose an alternative design for this baby rocking chair with modular components."}, which shifted the focus from a standard design to one featuring modular components.

Meanwhile, Midjourney enhanced the idea evolution stage by \textbf{promoting design concept iteration} and \textbf{detailing visual enhancements}. Specifically,  it facilitated rapid visualization and iteration of revised design concepts, enabling designers to swiftly explore and visualize modifications \textit{(P2-Combined Group)}. Additionally, it refined and detailed visual elements based on the same design theme, adding aesthetic intricacies that enriched the overall design \textit{(P5-Midjourney Group)}.

\section{Discussion}
\subsection{The role of Generative AI in human-AI collaboration in conceptual design}

\textbf{Generative AI expands designers’ solution exploration space and improves solution quality.} Based on the results, all three Generative AI-assisted groups extensively utilized the provided Generative AI models during the idea-generation stage, as illustrated in Figure~\ref{fig:sankey}~(a). This extensive use is likely due to the contextual solution generation capabilities of Generative AI, a notable advantage highlighted in previous research \citep{wu2024integrating, weisz2024design, lee2024and}. From the perspective of designers' evaluation, the assistance of Generative AI facilitates the exploration of a broader solution space. Specifically, with the support of Generative AI, scores for diversity, novelty, and the ability to trigger more ideas are all higher compared to those in the Human Group. From the perspective of design solution quality, experimental groups using Generative AI tools achieved higher mean scores across all five metrics examined in this study compared to the Human Group. Furthermore, overall scores for both the ChatGPT Group and Midjourney Group were found statistically significant differences compared to Human Group. This empirical evidence underscores the effectiveness of Generative AI in aiding novice designers during the conceptual design process. However, as this study aims to investigate ``How Generative AI supports humans in conceptual design", our primary focus is on the role of Generative AI in ``triggering more ideas". This focus may overlook the issue of design fixation potentially caused by Generative AI \citep{jansson1991design, wadinambiarachchi2024effects}, leaving room for future research to explore this further.

\textbf{With the assistance of Generative AI, human’s idea selection and evaluation stage was further triggered.} This finding stemmed from the post-interview data, where we asked participants to reflect on their design process from the perspective of conceptual design stages. When novice designers finish the design task on their own, the solution selection and evaluation stage may be overlooked (\textit{P2-Human Group}, \textit{P4-Human Group}). One possible explanation for this is that during the conceptual design process, designers often independently develop solutions starting from existing ideas (\textit{P2-Human Group: ``My strategy is that when I create this design, it was based on the existing possible problem with the baby and the stroller. This direct design process did not have an idea selection and evaluation process. It is a direct design and ignores the selection and evaluation processes."}). In this context, they primarily engage in autonomous concept development and find it challenging to step outside their established cognitive frameworks to effectively evaluate and select among different solutions. However, the pattern changes when designers collaborate with Generative AI —the involvement in the solution selection and evaluation stage becomes more pronounced. This suggests that Generative AI may enhance the solution selection and evaluation stage by prompting designers to critically assess and justify the outputs it generates. This interaction may help break cognitive biases and encourage a more thorough evaluation process (\textit{P2-Combined Group: ``I reviewed everything ChatGPT and Midjourney generated. Some evident flaws would be found. Following that, I also got some new ideas about solving the problem."}).

\textbf{Comparison between text-to-text and text-to-image models in initial conceptual design processes.} Although the experimental results indicated that Generative AI primarily assisted in the problem definition and idea generation stages, text-to-text models and text-to-image models played distinct roles in these two phases. Specifically, in the problem definition stage, ChatGPT was able to outline key points of product design and provide suggestions for innovative designs (\textit{P5-ChatGPT Group: ``For the first task, I only have a general idea and did not know the specific details. So I asked GPT-3.5 to tell me what the needed functions should be. In the second task, I don't really know about how to design musical bricks, and I command GPT-3.5 to tell me what the design of musical bricks commonly encompasses and which aspects I could make innovations in"}), owing to the extensive knowledge base and the capacity for a certain level of reasoning. On the contrary, while Midjourney could offer help in the problem definition stage, it requires users to input solution-oriented prompts, which necessitates the user having a preliminary idea about the design solution, as P2-Midjourney Group noted: \textit{P5-ChatGPT Group: ``Because it (Midjourney) relies on the initial keywords I provide. Without these keywords about design direction, it might deviate entirely from my intended idea"}. In essence, while Midjourney offers assistance in the problem definition stage, it is insufficient on its own. 

Although ChatGPT excelled in helping designers analyze individual design elements in the problem definition stage, the integrated design solutions generated from the filtered design points may confuse novice designers at idea generation stage. Conversely, Midjourney's advantage in visualization saves designers time in expressing ideas related to shape, texture, and color. For example, P1-ChatGPT Group inputted, \textit{Provide design ideas based on the elements I give you: ``a baby seat, appearance of Super Mario, blue and red as main colors”. Integrate the above design points and make it more complete and detailed."} However, ChatGPT’s response remained in the form of key points (such as theme and color scheme, shape and features, and fabric and materials). \textit{``It could not provide me an overview of the design solution"}, as expressed by P1-ChatGPT Group.

\subsection{Implications for future conceptual design support under Generative AI’s help}

\textbf{Workflow guidance and system integration should be carefully considered when combining text-to-text and text-to-image models.} In our experiment, the combination of ChatGPT and Midjourney did not yield a synergistic effect, both reflected in the participants' assessment of Generative AI tools and expert ratings results. Interview results suggest a possible explanation, as P5-Combined Group noted, \textit{``In the experiment, I primarily copied results from GPT-3.5 to Midjourney, but these models interpret my commands and produce results differently"}. This underscores how frustration could negatively impact the user experience with Generative AI, which might affect the outcome quality of human-AI co-creation solutions. Therefore, there is a necessity for methodologies and system designs that integrate the demands of various stages of conceptual design with the strengths of text-to-text and text-to-image models, respectively. Exploring better integration between these models could help leverage their combined potential.

\textbf{Explore the effect of image stimuli on stimulating designers' inspiration.} In previous research on Generative AI-enhanced conceptual design, the problem exploration stage was primarily enhanced by text-to-text models \citep{norheim2024challenges}, likely due to text being a fundamental mode of information expression. However, this study found that participants’ feedback indicated the highest novelty scores were achieved by Midjourney, and in expert ratings, the Midjourney Group also obtained higher mean scores than the ChatGPT Group. Although the text-to-text model leverages a big knowledge base to compensate for designers' limitations in knowledge and experience, human designers may overlook potentially important details due to the extensive textual information. Therefore, future system development could consider aiding designers in integrating information output by text-to-text models with visual design elements, or exploring the potential of visual search \citep{son2024genquery}, which could help designers relate the LLMs’ response to possible design solutions and enhance the role of image stimuli in inspiring designers’ creativity. 

\subsection{Limitations and future directions}
In this study, we focused exclusively on two representative generative models: a text-to-text model (ChatGPT) and a text-to-image model (Midjourney). Our decision regarding the specific choice of input and output modality was twofold. Firstly, the choice of text as the primary input modality was driven by its accessibility and familiarity, particularly for novice designers, facilitating easier expression of design intents. Secondly, for text and images as output modalities, they are regarded as the most commonly utilized data modalities in conceptual design, making them appropriate for our output modalities. 

Regarding continuous technical enhancements, on the one hand, more modalities can be incorporated to facilitate more flexible and naturalistic communication into the human-AI collaboration process, such as integrating voice, video, and sketches. By incorporating more modalities of Generative AI, researchers can more closely investigate the actual workflows of designers in experimental settings. On the other hand, with improvements in the generative models used in our research, such as GPT-4 and GPT-4o, researchers could explore two main directions in future work. One direction involves assessing their performances in processing multi-modal inputs. The other examines how these models perform in various types of design tasks, particularly those requiring more reasoning abilities, since previous research has revealed that as the complexity of tasks increases, the accuracy of the outputs generated by LLMs decreases \citep{khot2023decomposed}. We believe these new avenues for subsequent empirical research could significantly contribute to the ongoing refinement and application of various Generative AI technologies across different design scenarios.

On the other hand, this study investigated the differences between groups assisted by Generative AI and those who completed tasks independently, which aims to uncover how these general-purpose Generative AI tools enhance designers’ conceptual design processes compared to undertaking design tasks independently. Considering the specialized nature of design-specific tools, which are usually tailored to particular stages of conceptual design \citep{lee2024and}, future work could explore how workflow instructions and prompt engineering methods might affect the stages where Generative AI proves most beneficial.

For the choice of control group in this study, we selected Human Group for the purpose of comparing the differences and performances of human designers with and without the assistance of Generative AI. This approach helped us obtain some insightful findings and implications for future research. For example, with the assistance of Generative AI, the idea selection and evaluation stage was further triggered. Future work could include a comparative analysis with other traditional design support methods and tools, which would help provide a more comprehensive understanding of the value added by Generative AI.

Lastly, our experiment revealed that participants in the Combined Group, despite not being restricted on the order of tool usage, consistently used ChatGPT first, followed by Midjourney. Investigating the impact of the sequence of tool usage on experimental outcomes could provide valuable insights. Additionally,expanding our study to include a broader range of participants, such as more experienced designers, could help validate and extend our findings across different levels of expertise.

\section{Conclusion}
Our work aimed to investigate how Generative AI assists humans in the conceptual design process, especially for novice designers. Specifically, we conducted an experimental study involving 20 novice designers, assessing their performance with or without the help of text-to-text and text-to-image Generative AI models. The results revealed that Generative AI mainly assists humans in the initial stages of conceptual design, such as problem definition and concept generation, while the stages of idea selection and evaluation remains predominantly human-led. Despite the assistance of Generative AI, which improved participants’ feedback and expert ratings, the combination of text-to-text and text-to-image models did not exhibit an synergistic effect. Based on the findings, we discuss the role of Generative AI in human-AI collaboration and compare the efficacy of different models in design assistance. Ultimately, we propose several implications for enhancing the effectiveness and user-friendliness of human-AI collaboration in conceptual design. 

\section{Acknowledgements}
We thank all the participants for their time and the anonymous reviewers for their valuable comments. This research is supported by National Key R\&D Program of China (2022YFB3303304).

\bibliographystyle{agsm}  
\bibliography{ref}

\end{document}